\documentclass[12pt]{article}

\def\v{\vskip1em}

\def\ep{\epsilon}

\def\ra{\rightarrow}
\def\tint{{\textstyle\int}}

\def\d{\partial}
\def\dag{\dagger}

\def\b{\begin{eqnarray*}}  %takes no eqn numbers
\def\e{\end{eqnarray*}}    %takes no eqn numbers
\def\bn{\begin{eqnarray}}  %takes eqn numbers
\def\en{\end{eqnarray}}   %takes eqn numbers

\def\<{\langle}
\def\>{\rangle}

\def\no{\nonumber}

\def\{{\lbrace}

\def\d3{d^3\!x}

\def\b{\beta}

\def\}{\rbrace}
\begin{document} % , JQC Deserves, Requires, Merits

\title{The Particle in a Box\\ Warrants an Examination }
\author{John R. Klauder\footnote{klauder@ufl.edu} 
\\Department of Physics and Department of Mathematics  \\ %THE WORD "and" SEPARATES THE TWO %DEPARTMENTS
University of Florida,   %P.O. Box 118440\\
Gainesville, FL 32611-8440}
\date{}
\let\frak\cal
%bibliographystyle{

\maketitle 
\begin{abstract} 
The particle in a box is a simple model that has a classical Hamiltonian $H=p^2$ (using $2m=1$), with a limited  coordinate space,
 $-b<q<b$, where $0<b<\infty$. Using canonical quantization, this example has been fully studied thanks to its simplicity, and it is a common example for beginners to understand. Despite its repeated analysis, there is a feature that puts    the past  results into question. In addition to pointing out the quantization issue, the procedures of affine quantization can lead to a proper quantization that nesaeccsrily  points toward more complicated eigenfunctions and eigenvalues, which deserve to be solved.
 \end{abstract}
 
 \section{A Standard Quantization of \\the Particle in a Box} 
 The basic operators of canonical quantization (CQ), 
 namely $P$ and $Q$, have appropriate  representations that range over entire real values, which leads both operators to being self-adjoint, i.e., $P^\dag=P$ and $Q^\dag =Q$. These operators admit a Schr\"odinger representation, e.g.., $P\ra -i\hbar (d/dx)$ and $Q\ra x$, with $-\infty<x<\infty$.
 
 Since our story takes place entirely in the interval $-b<x<b$, we seek the solutions of 
  \bn -\hbar^2 (d^2 \phi(x) /dx^2) \;= E\;\phi(x)\en 
    with the requirement $\phi(-b)=\phi(b)=0$ to match the fact that $\phi(x)=0$ for all  of  $x\leq-b$ and $x \geq b$, thanks to `implicit infinite walls'. This connection at $x=\pm b$ ensures that $\phi(x)$ will be continuous. Clearly, the $\cos$ and $\sin$ functions fit the requirement, like $\phi_{n}(x)= \cos(n\pi x/2b)$, for $n=1,3,5,...$, and
     $\phi_n(x)=\sin(n\pi x/2b)$, for$n=2,4,6,...$.
    The eigenvalues then become $E_n = n^2\pi^2/4b^2$, now for $n=1,2,3,4,5...$. Evidently, 
    two derivatives of either $\sin$ or $\cos$ just   leads to a simple factor times the original term.
    The norm of the eigenfunctions becomes 
  $\tint_{-b}^b
    |\cos(n\pi x/2b)|^2\;dx<\infty$
    for $n=1,3,5,...$, and 
    $\int_{-b}^b 
    |\sin(n\pi x/2b)|^2\;dx<\infty$
    for $n=2,4,6,...$. Hence, the Hilbert space is composed of arbitrary sums of these eigenfunctions that lead to finite normalizations. 
    
    Clearly, this first section points toward a bona-fide quantization of a particle in a box that has been uniformly accepted; for a standard  analysis, see \cite{wpp}.

    \section{A Common Review of Derivatives}
    \subsection{A toy example of the problem}
  The feature that is of concern deals with the proper derivatives of the prospective eigenfunctions. This feature can be seen in a simple example.
  
  Consider the continuous function $f(x)$, where $-1<x<1$, which is defined as $f(x)=0$ for $-1\leq x \leq0$, and $f(x)=x$ for $0\leq x\leq 1$.\footnote{Picture a flat floor under a flat door that leans upward from the floor at 45 degrees.} A simple review of derivatives shows that  $f'(x)= \lim_{\ep\ra0}
  [f(x+\ep) -f(x-\ep)]/2\ep$, which, for our example, leads to $f'(x)=0$ for $x<0$ and $f'(x)=1$
  for $x>0$. In addition, we find that $f'(0)=1/2$.
  This result has led to $f'(x)$ becoming a discontinuous function. The next step is defining 
  $f''(x)=\lim_{\ep\ra0} [f'(x+\ep)-f'(x-\ep)]/2\ep $. The result is that $f''(x)=0$ for $x<0$ and  $f''(x)=0$ for $x>0$. In addition, it follows that
  $f''(0)=\lim_{\ep\ra0} [f'(\ep) - f'(-\ep)]/2\ep] =\infty$. If we imagine that  $2\ep\ra dx$, then $f''(x)=\delta(x)$, which is Dirac's delta function,   formally chosen so that $\tint_{-1}^1 \delta(x)\,dx=1$. Summarizing, we have found that $f''(x) = \delta (x)$. This result has been because $f'(x)$ is a discontinuous function. Let us add that while 
 $\tint_{-1}^1 |f''(x)|\; dx<\infty$, it follows that 
 $\tint_{-1}^1 |f''(x)|^2\; dx=\infty$, which is the reason to reject $f''(x)$ in any   Hilbert space.
 
 \subsection{Application to a particle in a box}
 The proposed eigenfunction ground state of the particle in a box starts with 
 \bn &&\hskip-2.5em \phi_1(x)=0 \;\;\;(for \; x\leq-b) \no \\
 && =\cos(\pi x/2b)  \;\;\; (for\; -b\leq x \leq b)\no \\ 
   && =0  \;\;\; (for \;b\leq  x). \en
   However, two derivatives of the eigenfunction ground state, i.e., $\phi_1(x)=\cos(\pi x/2b)$, lead to 
 \bn &&\hskip-2.6em \phi_1''(x)=0 \;\;\;(for \; x<-b) \no \\
 &&= (\pi^2/8b^2) \,\delta(x+b) \no\\
 && =-(\pi^2/4b^2)\cos(\pi x/2b)  \;\;\; (for\; -b < x  <b)\no \\ 
  &&=-(\pi^2/8b^2) \,\delta(x-b)\no\\
   && =0  \;\;\; (for \;b<  x). \en
   This equation does not match the desired form of $-\hbar^2 \phi_1''(x)= E_1\, \phi_1(x)$, and moreover, 
  $\tint_{-\infty}^\infty |\phi_1''(x)|^2\;dx =\infty$, 
   which can not be accepted by any Hilbert space.
   
    While we have only considered the standard ground state, every proposed eigenfunction would have a similar story and {\it none of them could  be a proper eigenfunction with a finite eigenvalue}, e.g., $-\hbar^2\phi_n''(x)= E_n \;\phi_n(x)$ with $|E_n|<\infty$. Nor would they belong to the usual Hilbert space due to having $\delta(x\pm b)$ terms in their second derivative.
    In the author's opinion, this behavior fails the conventual quantization of a particle in a box,
    as it was reviewed in Sec.~1.

    In the next section, we will propose a valid quantization of the particle in a box using an alternate quantization procedure. 
    
  \subsection{Half of the expected eigenfunctions}
  The sentence, taken from Sec.~1,
  ``This connection at $x\pm b$ ensures that $\phi(x)$ will be continuous. Clearly, the $\cos$ and $\sin$ functions fit the requirement, like $\phi_{n}(x)= \cos(n\pi x/2b)$, for $n=1,3,5,...$, and
     $\phi_n(x)=\sin(n\pi x/2b)$, for $n=2,4,6,...$,'' makes it clear that only  half of the sin and cos terms have been accepted. This means that the standard box treatment leads to only half of the  set of eigenfunctions that were  presumed.

    \section{A Valid Quantization of \\the Particle in a Box}
   Instead of CQ, we use affine quantization (AQ)
   in order to produce a valid quantization of the  particle in a box. Fortunately, AQ has been {\it designed} to deal with reduced coordinate spaces. 
   
   \subsection{A brief review of affine quantization}
   Some classical problems require reduced coordinates, like the half-harmonic oscillator with the classical Hamiltonian $H=(p^2+q^2)/2$ provided that $q>0$. In so doing, $P^\dag\neq P$, and we seek substitutes for $p$ and $P$. We choose the dilation variable $d=pq$ and $q$ as the new variables. To ensure their independence, we eliminate $q=0$, and then discard all $q<0$, keeping all $q>0$.\footnote{A real-life example, is cutting a long thread into two parts and only keeping one of them.} The basic affine quantum operators are now $D=(P^\dag Q+QP)/2$ and $Q>0$. Coherent states, used to connect classical and quantum realms [2], have established
   that with special classical variables we find that $H'(d,q)\ra H'(D,Q)$. This implies that the classical Hamiltonian can now be $H'=(d^2/q^2 +q^2)/2$, along with $q>0$, and the affine quantum Hamiltonian for this model then becomes 
   \bn && H'=(D Q^{-2} D + Q^2)/2
       =[P^2+ (3/4)\hbar^2/Q^2 +Q^2]/2 \:.\en
  Happily, the `3/4' term ensures that both $P^\dag$ and $(P^\dag)^2$  {\it act} like $P$ and $P^2$ in this equation. 
  
  This example has proved its validity in several papers that have exposed the eigenfunctions and eigenvalues for the half-harmonic oscillator [3, 4, 5]. This favorable result for the eigenfunctions being of the form
  $\phi_n(x)=x^{3/2} (polynomial)_n e^{-x^2/2\hbar}$, with $n=0,1,2,...$, a fact that the first derivative remains a completely continuous function. To see what the $x^{3/2}$ factor can do for the relevant two derivatives, just examine the simple equation $[-\hbar^2 \,(d^2/dx^2) + (3/4) \hbar^2 /x^2]\; x^{3/2} = ?$.

  \subsection{Affine quantization of the particle in a box}
  The example of the last section removed $q=0$ from the coordinate space. Now we remove two points, namely  $q=-b$ and $q=b$. That leaves three different spaces, and we keep only the middle part, namely, where $-b<q<b$. We use a suitable dilation variable, namely, $d'=p(b^2-q^2)$, along with $q$, which  is restricted by $-b<q<b$. Next, we offer $D' =[P^\dag (b^2-Q^2)+(b^2-Q^2) P]/2$, and $-b<Q<b$. 
  
  Using the new affine variables, the classical Hamiltonian is  $p^2 = d'(b^2-q^2)^{-2} d'$, which is promoted to
   \bn H' = D' (b^2-Q^2)^{-2} D' = 
   P^2 +\hbar^2[2Q^2+b^2]/[(b^2-Q^2)^2]\;. \en
 This $\hbar$-factor has been selected from a general study of various dilation operators in [6].
    
   It is noteworthy to examine the last equation when $Q$ accepts a Schr\"odinger representation, i.e., $Q=x$, and $x$ is extremely close to either $\pm b$. In that case,
 $[2x^2+b^2]/[(b^2-x^2)^2] \simeq (3/4)/(b\pm x)^2$, a relation that closely resembles the very strong properties from the present model with those of the former  model. Using this fact leads to the suggestion that part of any eigenfunction of (5) is likely to be $\psi(x)= (b^2-x^2)^{3/2} (remainder)$.\footnote{A very different use of (5) is to {\it accept} the outside space, $|x|>b$, and {\it reject} $|x|<b$, which then it becomes an `anti-box'. Now this system has a similarity to a toy `black hole'. It may happen that particles  could pile  up close to an `end of space', while being attracted by a simple,  `gravity-like', pull of a potential such as  $V(x)= W/|x|$. If you choose AQ, the $\hbar$-term in (5) could prevent the particles from falling   `out of space'.}

Efforts to find eigenvalues and eigenfunctions 
for (5) are open to help shed further information on this effort to find a proper quantization of the particle in a box. 

\section{Conclusion}
Affine quantization has also been able to help with other problems as well. Recent efforts have been focused on quantum field theory [7], and the intro- duction of an affine path integral quantization of gravity [8]. Another article features a direct comparison of CQ and AQ regarding the half-harmonic oscillator with its important implications for gravity and field theory [9].\v

Thanks are given to B.-S. Skagerstam, S. Shabanov, J. Govaerts, and A. Kempf for helpful comments of earlier versions of this paper.\v

{\bf Note:} The author has no conflicts to disclose.

\end{document}